\documentclass[superscriptaddress,showkeys,amsmath,amssymb,pra,twocolumn]{revtex4}

\usepackage{graphicx,amssymb}
\usepackage{dcolumn}
\usepackage{bm}
\usepackage{subfig}
\usepackage{relsize}
\usepackage{nccmath}
\usepackage{color}

\newcommand{\beq}{\begin{equation}}
\newcommand{\eeq}{\end{equation}}
\newcommand{\bea}{\begin{eqnarray}}
\newcommand{\eea}{\end{eqnarray}}
\newcommand{\e}[1]{\times 10^{#1}}

\begin{document}
\title{Effects of Interactions and Temperature in Disordered Ultra-Cold Bose Gases}
\date{\today}
\author{C.~P.~J.~Adolphs}
\affiliation{The Jack Dodd Centre for Quantum Technologies, Department of Physics, University of Otago, Dunedin, New Zealand}
\author{J.~Towers}
\affiliation{The Jack Dodd Centre for Quantum Technologies, Department of Physics, University of Otago, Dunedin, New Zealand}
\author{M.~Piraud}
\affiliation{Laboratoire Charles Fabry de l'Institut d'Optique, CNRS et
Universit\'e Paris-Sud, Campus Polytechnique, RD 128, F-91127 Palaiseau cedex, France}
\affiliation{The Jack Dodd Centre for Quantum Technologies, Department of Physics, University of Otago, Dunedin, New Zealand}
\author{K.~V.~Krutitsky}
\affiliation{{Fakult\"at f\"ur Physik} der Universit\"at Duisburg-Essen, Campus Duisburg, Duisburg, Germany}
\author{D.~A.~W. Hutchinson}
\affiliation{The Jack Dodd Centre for Quantum Technologies, Department of Physics, University of Otago, Dunedin, New Zealand}
\begin{abstract}
We simulate ultra-cold interacting Bosons in quasi-one-dimensional, incommensurate optical lattices. In the tight-binding limit, these  lattices have pseudo-random on-site energies and thus can potentially lead to Anderson localization. We explore the parameter regimes that lead to Anderson localization and investigate the role of repulsive interactions, harmonic confinement and finite temperature. We find that interactions can obscure the exponential localization characteristic of Anderson localization, thus impeding the direct observation of this phenomenon when interactions are present.
\end{abstract}
\maketitle
Atoms in optical lattices provide a great tool to study many theoretical models from solid-state physics as they allow for unique control of normally unmodifiable parameters. Disorder can be introduced and controlled by external optical potentials \cite{Lewenstein,Fallani} and interactions can be tuned by a Feshbach resonance \cite{Vogels}. While being intrinsic to a solid-state system, disorder can easily be imposed onto and modified in an optical lattice. Disorder in a quantum mechanical system leads to various interesting effects such as Anderson localization, which describes the localization of matter waves in a random potential due to destructive interference from randomly scattered waves \cite{Phil}. Several experimental groups have studied localization effects in ultra-cold Bose gases in disordered optical lattices. To create the disorder, either a laser speckle potential \cite{Sanchez} or a second optical lattice  incommensurate with the original lattice is superimposed onto it \cite{Lye,Schulte}. We extend mean field \cite{Pethick} treatments and their finite temperature generalizations \cite{Griffin,HZG,JPhysB}, which  have been extensively developed for the dilute Bose gas, including their application to the optical lattice \cite{Wild,Rey}, to include disorder.

We study a Bose gas in a quasi-one-dimensional disordered optical lattice. A tight-binding model leads to a Bose-Hubbard Hamiltonian \cite{Wild} which is further approximated and numerically solved. We study the effects of different disorder and interaction strengths on the condensate density, the condensate and superfluid fractions, the lowest excitation energy and the band structure. We use the same setup as Schulte {\it et al.} \cite{Schulte}, but also treat a harmonically confined Bose gas at finite temperature.

\section{Model}

\subsection{The system}

The model we use for our simulations is taken from \cite{Wild} and references therein. Tight radial confinement of a Bose gas leads to a quasi-one dimensional problem while scattering remains three-dimensional. The lattice potential is  $V_L(x) = V_{0} \cos^{2} (2 \pi x/\lambda_L)$, where $V_{0}$ is the lattice depth and $\lambda_L$ is the lattice wavelength. The recoil energy is defined as  $E_R = \hbar^{2} {k_{L}}^{2}/2 m$, where $k_{L}$ is the lattice wave-number.
The disorder potential is of the form
\beq
V_{dis}(x)
=
V_{dis}^{0}
\left[
    \cos^{2}
    \left(
        2\pi\alpha x/\lambda_L
    \right)
    +
    \cos^{2}
    \left(
        2\pi \beta x/\lambda_L
    \right)
\right]
\;,
\eeq
where $\alpha$ and $\beta$ are the incommensurate ratios between the primary and secondary lattice wavelengths. For irrational values $\alpha$ and $\beta$, the resulting potential is pseudo-random. In experiment, however, the wavelength ratios are always rational, so the potential exhibits an additional level of quasi-periodicity. If $\alpha$ and $\beta$ are ratios of large coprime integers, however, the period of the potential will be larger than the system size.

\subsection{The Bose-Hubbard Model}

The wavefunction is expanded in Wannier functions localized at each lattice point of the lattice potential. This tight-binding approximation leads to the Bose-Hubbard Hamiltonian for atoms in a one-dimensional optical lattice of $N$ sites,
\beq
\hat{H} = \sum_{i=1}^{N}{\hat{n}_{i} \epsilon_{i}} - \sum_{i=1}^{N-1}{J_{i,i+1} (\hat{a}_{i+1}^{\dagger} \hat{a}_{i} + \hat{a}_{i}^{\dagger} \hat{a}_{i+1})} + \frac{U}{2} \sum_{i=1}^{N}{\hat{n}_{i}(\hat{n}_{i}-1)},
\eeq
with tunneling energies $J_{i,i+1}$, on-site interaction potential $U$, and on-site energy $\epsilon_{i}$, which contains both the trapping potential and the disorder potential. These parameters are determined by the lattice and disorder potential and are obtained by taking integrals over the Wannier functions \cite{Morsch}. The resulting expressions in the Gaussian approximation
\footnote{In this approximation, the Wannier functions in the expectation-value integrals are replaced by Gaussians. This gives accurate results for the on-site energies while underestimating the tunneling energy. Since results are given in terms of the tunneling energy J, this is of no concern if one keeps in mind that a given value for J does not exactly correspond to the given lattice geometry.}
are
\beq \label{eqn:U}
U = 2 \sqrt{\frac{2}{\pi}} \frac{\hbar^{2}}{m a_{\perp}^{2}} k_{L} a_{s} \left(\frac{V_{0}}{E_{R}}\right)^{1/4},
\eeq
\beq \label{eqn:J}
J = V_{0} e^{-(\pi/2)^{2} \sqrt{\frac{V_{0}}{E_{R}}}} \left[ \left(\frac{\pi}{2}\right) - 1\right],
\eeq
where $a_{\perp} = \sqrt{\frac{2 \hbar}{m \omega_{\perp}}}$, $a_{s}$ is the scattering length of the atoms and $\omega_{\perp}$ the radial (perpendicular) trapping frequency.\\
The disorder potential leads to pseudo-random shifts in the on-site energies and the tunneling energies. The on-site energy for one secondary lattice with wavelength ratio $\alpha$ to the primary lattice wavelength is
\beq
\epsilon_{i}
=
\epsilon_0
\cos
\left[
    \pi \alpha (2i - 1)
\right]
\;,\quad
\epsilon_0
=
\frac{V_{dis}^{0}}{2}
e^{
    -\pi^{2}
    \sqrt{\frac{E_{R}}{V_{0}}}
  }
\;,
\eeq
for two secondary lattices with ratios $\alpha$ and $\beta$, the on-site energy is just the sum of the above term for each of the two ratios. Our simulations show that the site-dependence of the tunneling rate has a negligible effect on the condensate and we thus omit it here for simplicity, such that $J_{i,i+1} = J$ for any $i$.

\subsection{The Hartree-Fock-Bogoliubov Formalism}

When a macroscopic occupation of the ground state is assumed, the Bose annihilation operator can be written as $\hat{a}_{i} = (z_{i} + \hat{\delta}_{i}) e^{-i \mu t / \hbar}$, where $z_{i}$ is a complex mean-field part and $\hat{\delta}_{i}$ the fluctuation operator \cite{Wild}. The condensate density is given by $n_{c,i} = |z_{i}|^{2}$ and the non-condensate density by $\tilde{n}_i = \langle \hat{\delta}_{i}^{\dagger} \hat{\delta}_{i} \rangle$.\\
From this, we can construct a discretized, generalized Gross-Pitaevskii equation
\beq
\mu z_{i} = \epsilon_{i} z_{i} - J(z_{i+1} + z_{i-1}) + U(n_{c,i} + 2 \tilde{n}_{i} z_{i}),
\eeq
and Bogoliubov-de Gennes equations 		 
\bea
\hbar \omega_{q} u_{i}^{q} &= [2 U (n_{c,i}+\tilde{n}_{i}-\mu+\epsilon_{i}] - J [u_{i+1}^{q}+u_{i-1}^{q}] - U {z_{i}}^{2} v_{i}^{q}\nonumber\\
-\hbar \omega_{q} v_{i}^{q} &= [2 U (n_{c,i}+\tilde{n}_{i}-\mu+\epsilon_{i}] - J [v_{i+1}^{q}+v_{i-1}^{q}] - U {z_{i}}^{2} v_{i}^{q}\nonumber
\eea
with $\tilde{n}_{i} = \sum_{q}{\left[|v_{i}^{q}|^{2}+(|u_{i}^{q}|^{2}+|v_{i}^{q}|^{2}) N_{BE}(\hbar \omega) \right]}$, where $u_{i}^{q}$ and $v_{i}^{q}$ are the Bogoliubov quasi-particle amplitudes and $N_{BE}$ is the Bose-Einstein distribution. This is a lattice formulation\cite{Wild,Rey} of the finite-temperature Hartree-Fock-Bogoliubov \cite{Griffin,HZG,JPhysB} formalism. The equations are then solved self-consistently.

\section{Results}

Here we show results for total atom number $N = 500$ atoms of \textsuperscript{87}Rb in a system with $100$ lattice sites in the experimental configuration of \cite{Schulte}, where $\alpha = 55/64$ and $\beta = 165/212$.\\
The strength of inter-particle interactions can be tuned by a Feshbach resonance \cite{Vogels} and, for finer control, by the radial confinement.
Here we define $\Delta = \epsilon_0/J$ and $V_{eff} = U/J$
so that we need not be concerned with evaluating the integrals from experimental parameters.

\subsection{Homogeneous Trap at Zero Temperature}

To isolate the effects of disorder and interactions, the first simulations are performed without a harmonic trapping potential. We refer to this as the homogeneous case albeit dealing with a finite size system with hard wall boundary conditions.
\begin{figure}[h!]
 \centering{
 \subfloat[$\Delta = 0$]{
  \includegraphics[width=0.2\textwidth]{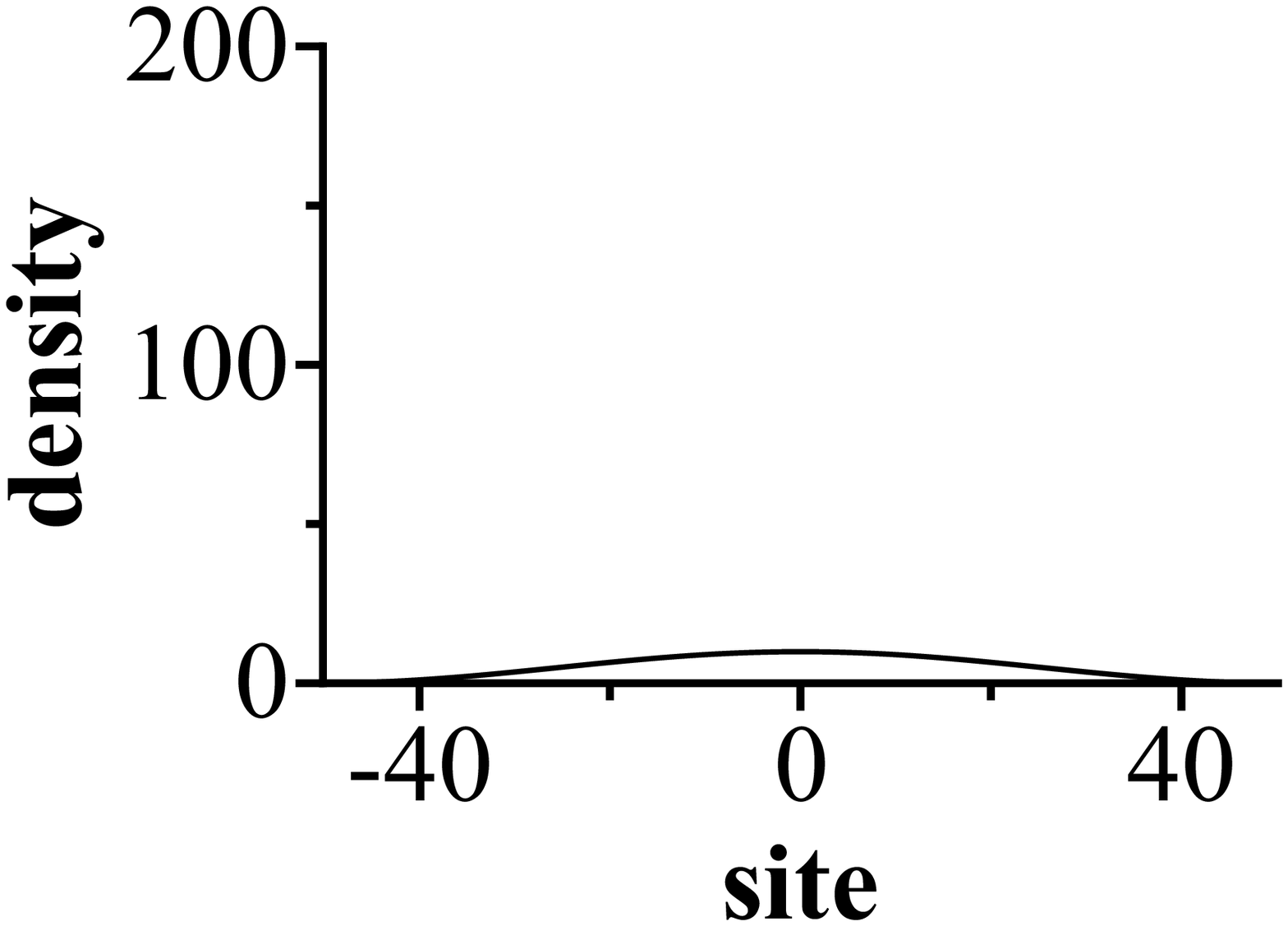}}
 \subfloat[$\Delta = 0.68$]{
  \includegraphics[width=0.2\textwidth]{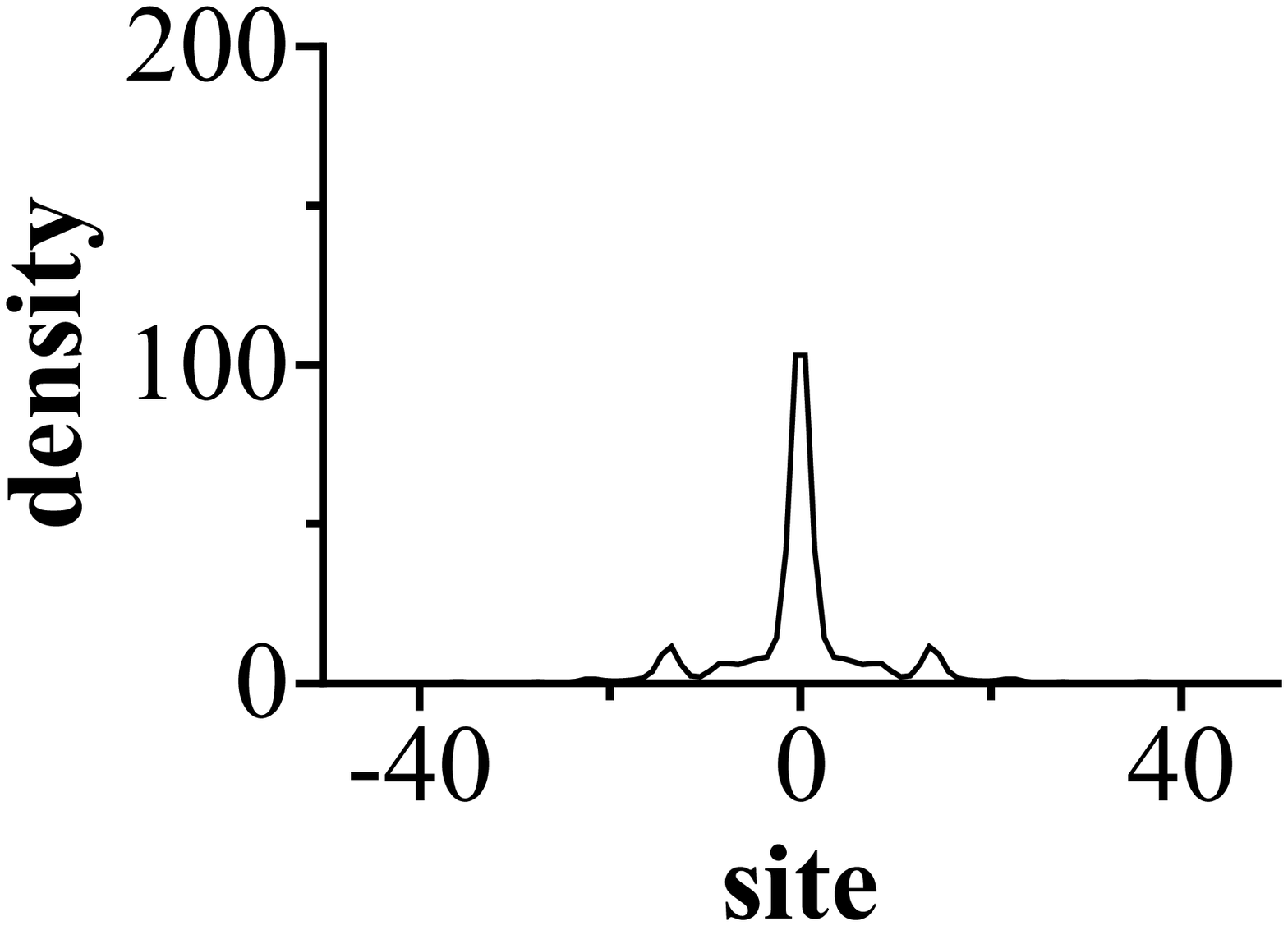}}

 \subfloat[$\Delta = 1.13$]{
  \includegraphics[width=0.2\textwidth]{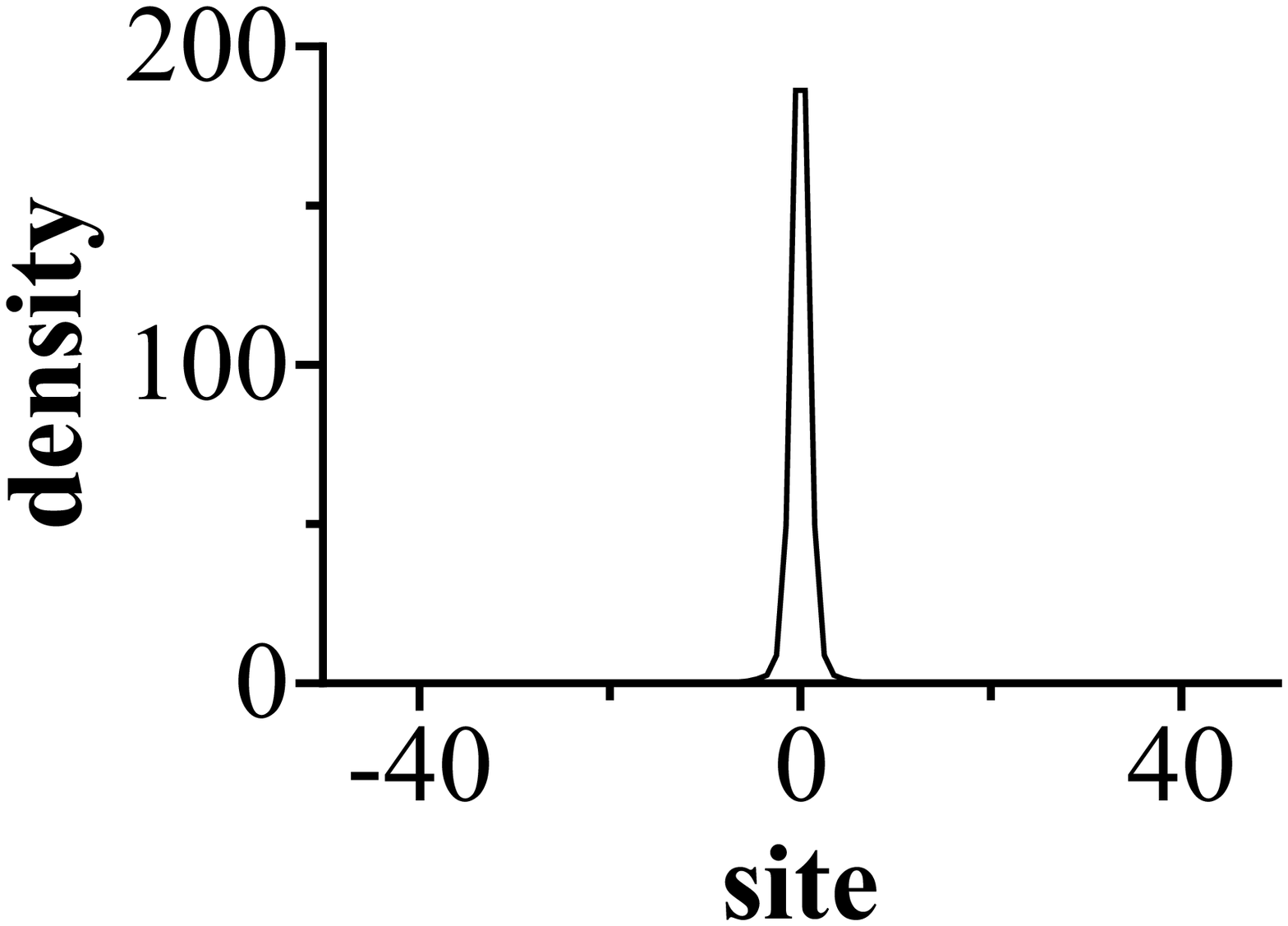}}
 \subfloat[$\Delta = 1.13$]{
  \includegraphics[width=0.2\textwidth]{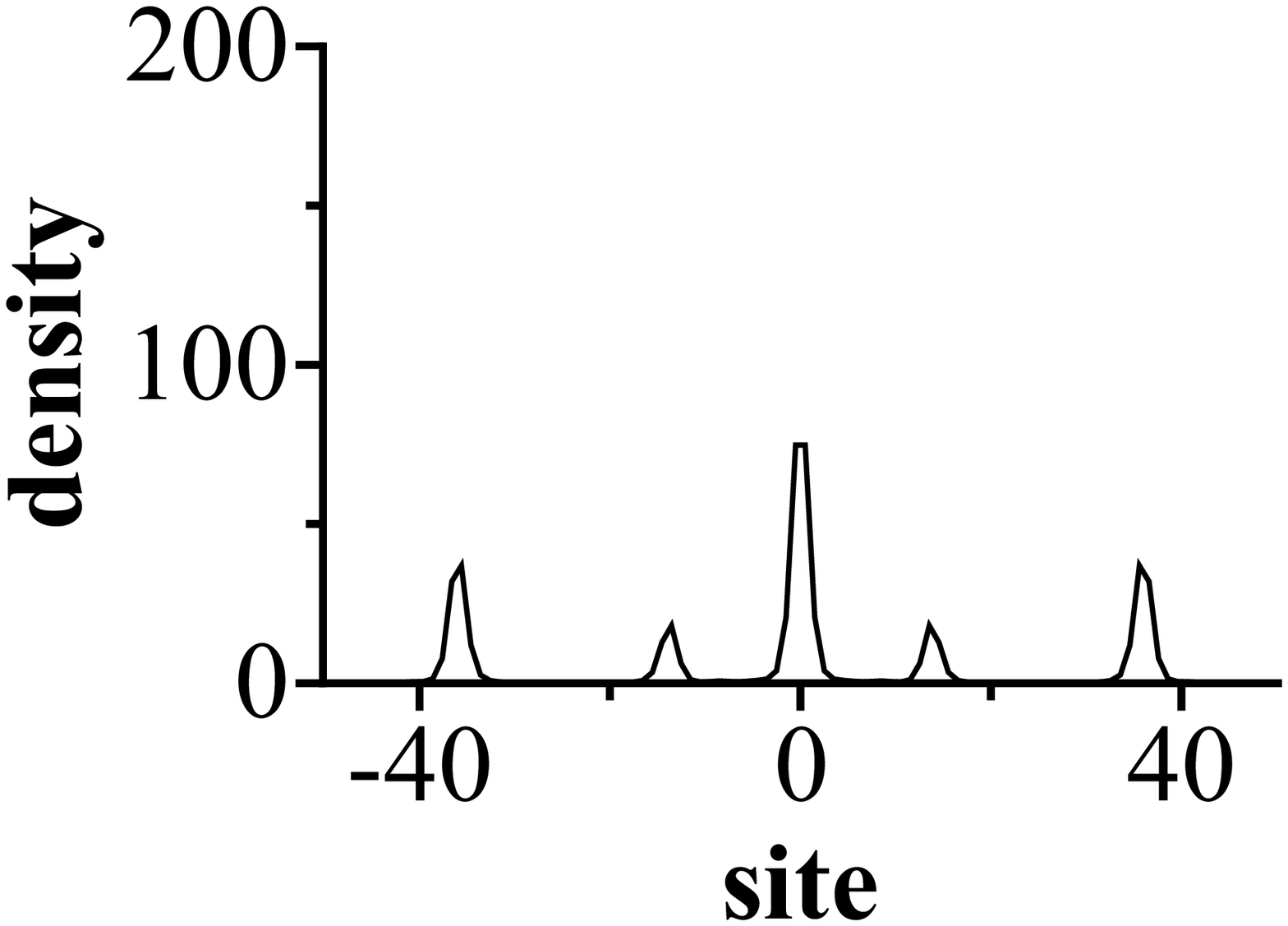}}
  }
 \caption{\label{fig:01}Condensate density for various parameter values. \mbox{(a-c) $V_{eff} = 10^{-8}$.} \mbox{(d) $V_{eff} = 6\e{-4}$.}}
\end{figure}
\begin{figure}[h!]
 \centering{
 \subfloat[$\Delta = 0$.]{
  \includegraphics[width=0.2\textwidth]{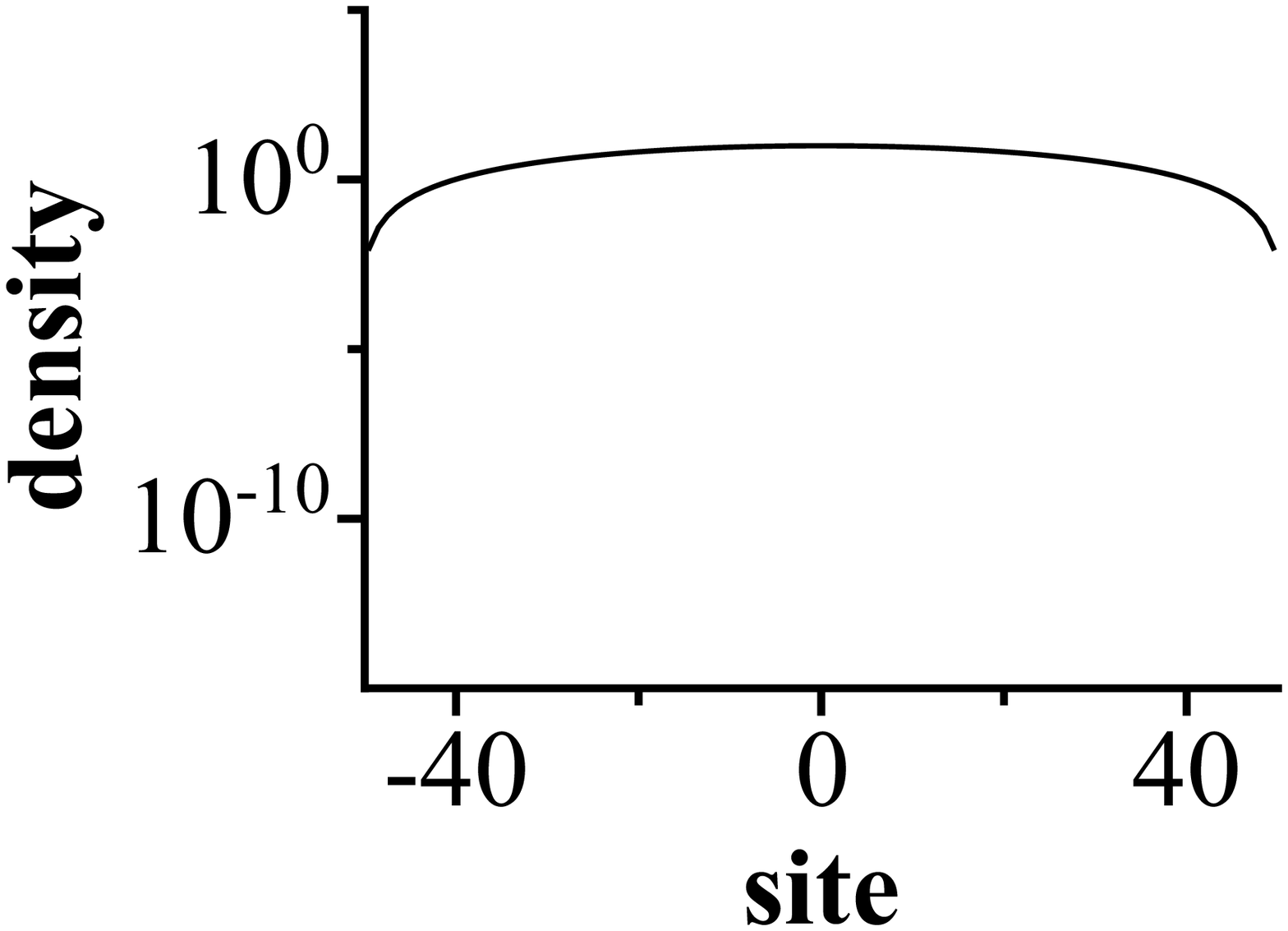}}
 \subfloat[$\Delta = 0.68$.]{
  \includegraphics[width=0.2\textwidth]{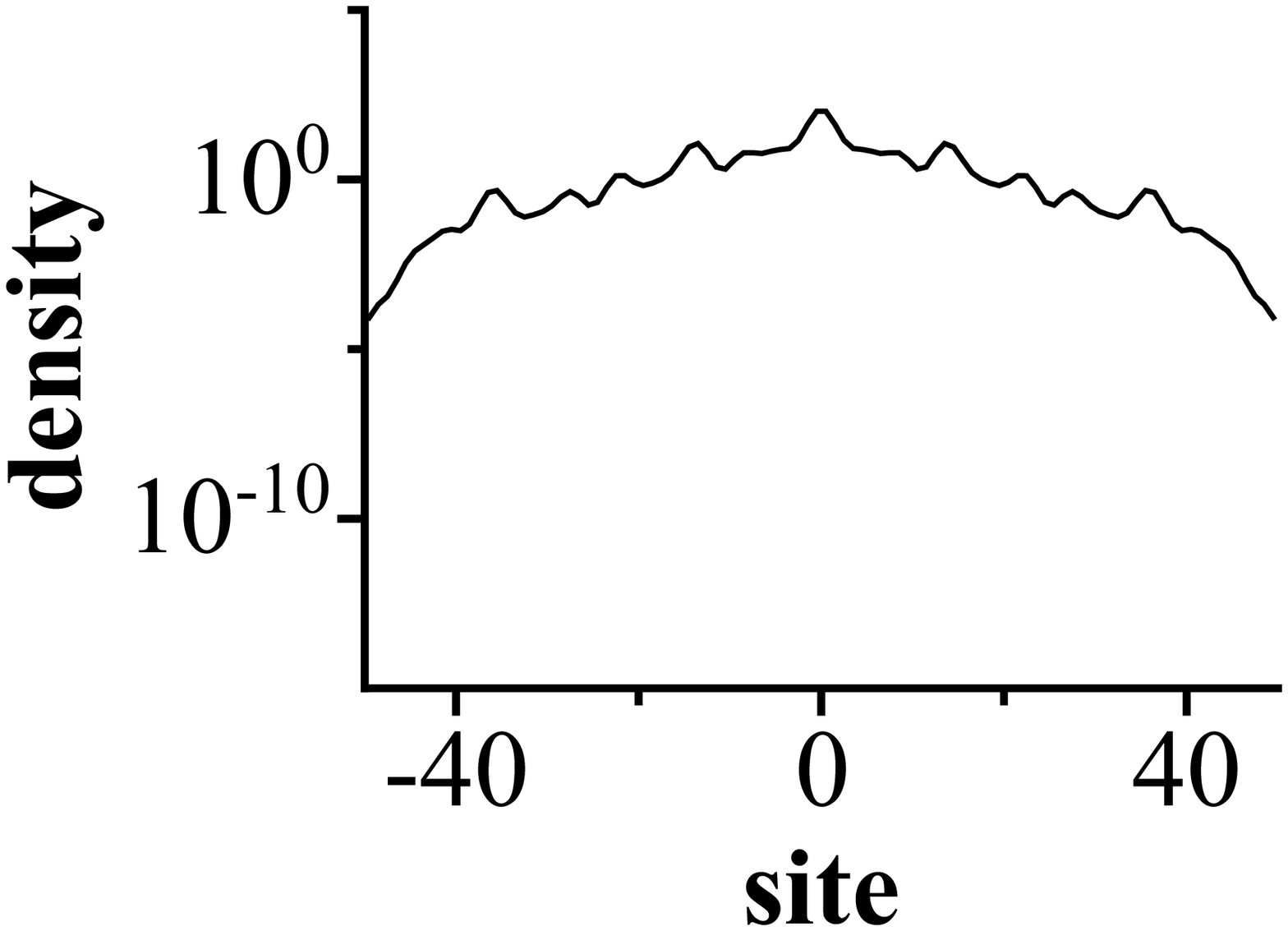}}

 \subfloat[$\Delta = 1.13$.]{
  \includegraphics[width=0.2\textwidth]{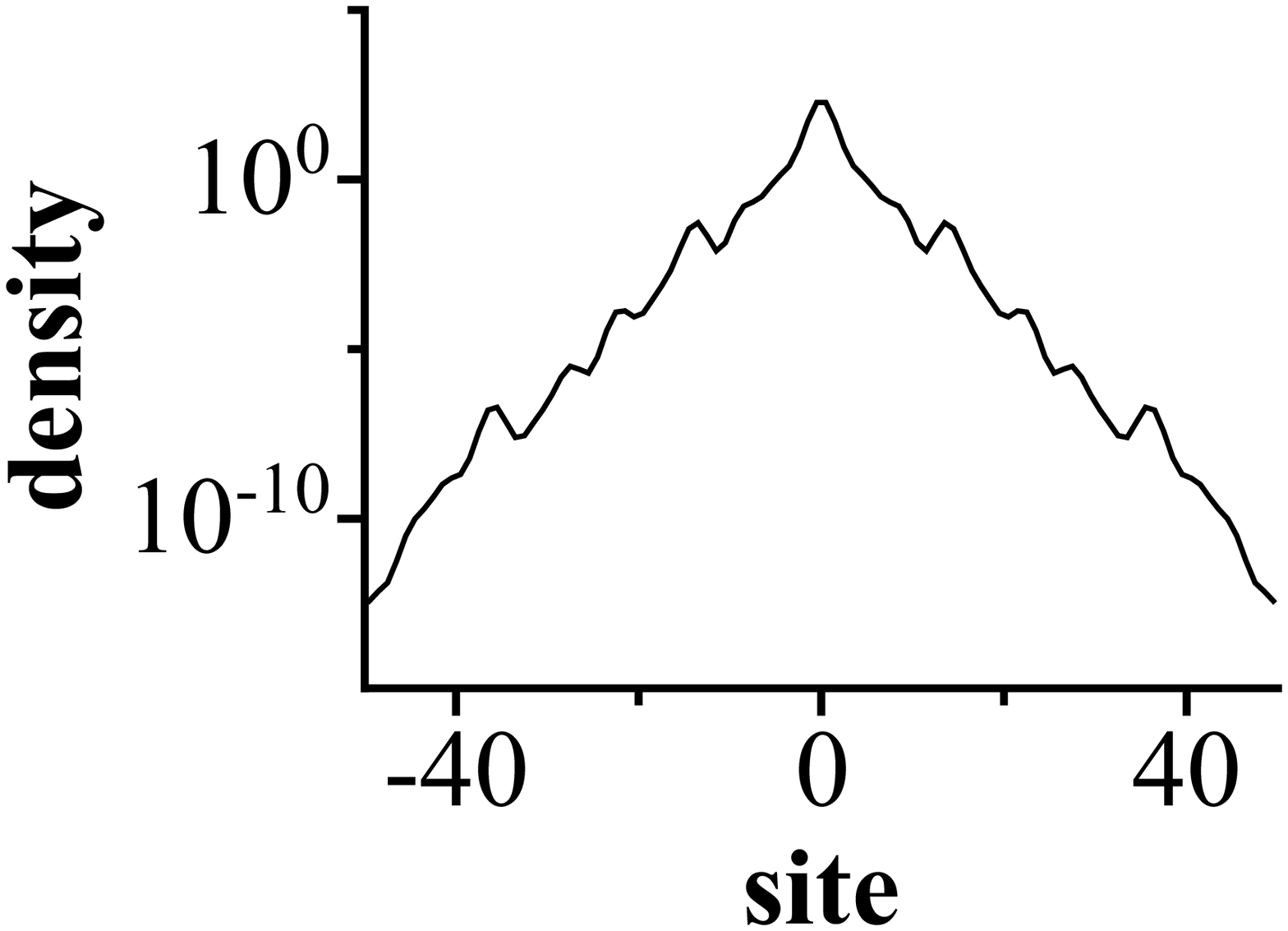}\label{sfig:02c}}
 \subfloat[$\Delta = 1.13$.]{
  \includegraphics[width=0.2\textwidth]{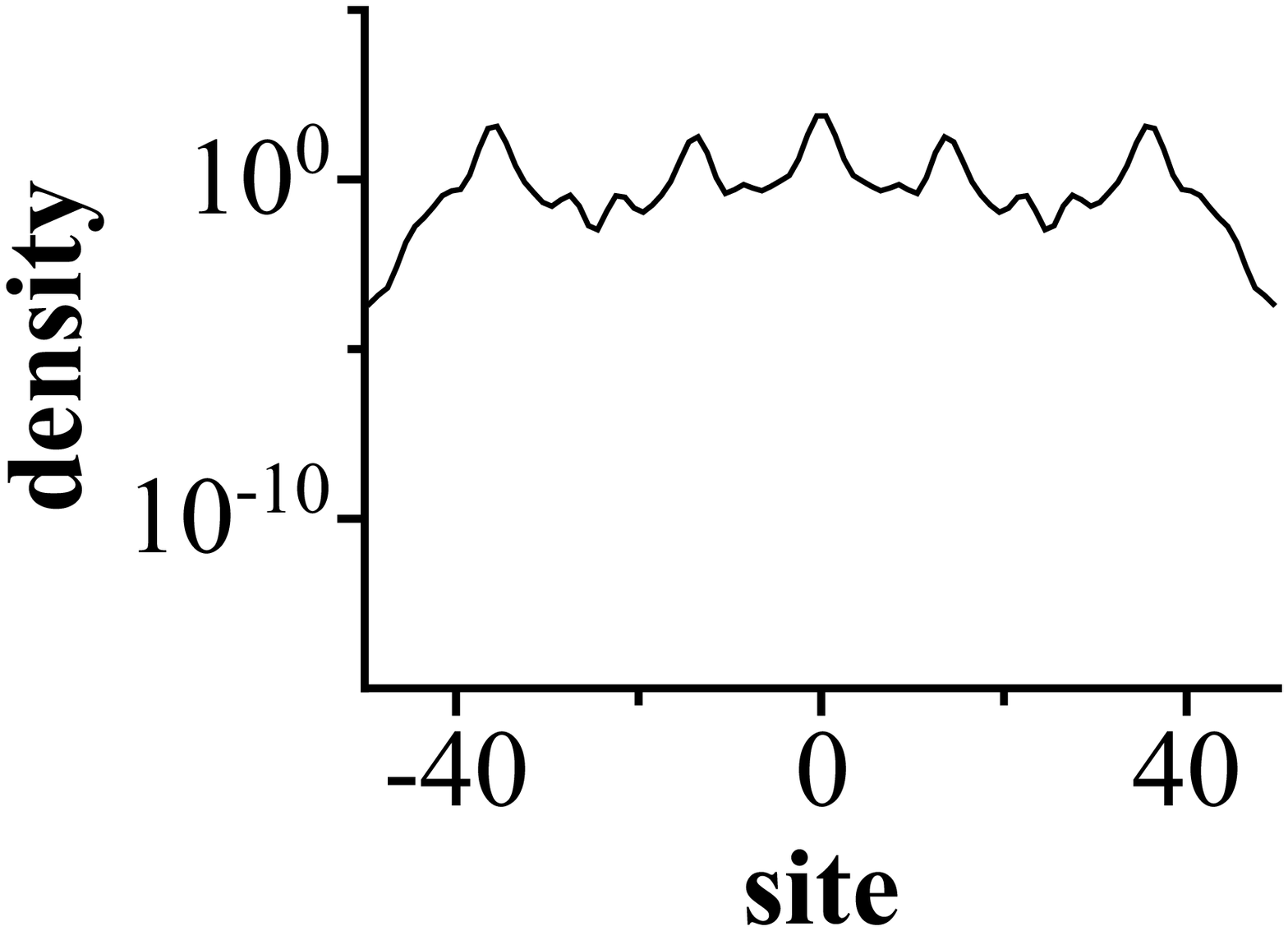}\label{sfig:02d}}
  }
 \caption{\label{fig:02}Condensate density (logarithmic) for various parameter values. \mbox{(a-c) $V_{eff} = 10^{-8}$. (d) $V_{eff} = 6\e{-4}$.}}
\end{figure}
Fig. \ref{fig:01} and \ref{fig:02} show the condensate density for different parameters in linear and logarithmic scale. The first panel shows the weakly interacting, non-disordered case, which is just the ground-state of a single particle in a box. Increased disorder then leads to a narrowly peaked density. The logarithmic plot in Fig. \ref{sfig:02c} indicates an exponential decay of the condensate density, a signature of Anderson localization. The Andr\'e-Aubry model \cite{Aubry}, which describes non-interacting Bosons in two incommensurate lattices, predicts a transition from extended to localized states for $\Delta = 2$. Here we have two secondary lattices instead of one, so the transition occurs for $\Delta = 1$. In the presence of interactions, the condensate fragments into multiple peaks. Each single peak remains exponentially localized (Fig. \ref{sfig:02d}), but limited spatial resolution in experiments might be unable to distinguish them.

\begin{figure}[h!]
 \includegraphics[width=0.45\textwidth]{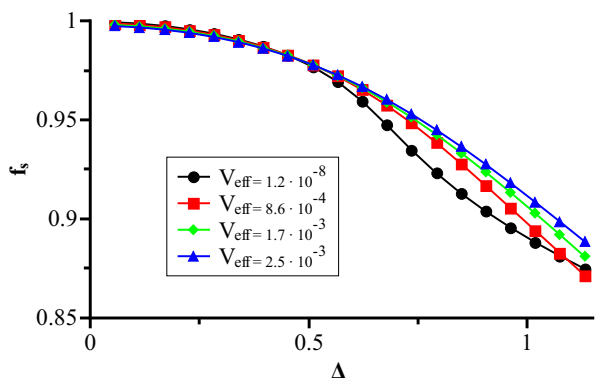}
 \caption{\label{fig:03}Superfluid fraction $f_{s}$ vs. $\Delta$ at various interaction strengths.}
\end{figure}

We find that increasing the disorder strength continuously lowers the superfluid fraction, underlining the localizing effect of disorder (Fig. \ref{fig:03}). The decrease is not dramatic as even for strong disorder the superfluid fraction remains above $0.85$. The long-range phase coherence implied by this may well be an artifact of our mean field technique, which is not strictly valid in the localized extreme. The technique should, however, be reliable up to this point and is thus a valid predictor of any transition.

In the case of weak disorder, the superfluid fraction decreases slightly with increased interaction strength, which is explained by repulsive interactions inhibiting superfluidity. In contrast, for increased disorder the superßuid fraction increases with higher $V_{eff}$ . This follows from the interplay between interactions and disorder. The disorder localizes the condensate, thereby reducing the superßuid fraction. The interactions delocalize the condensate and thus attenuate this effect. We note that for very large interaction strengths, the superfluid fraction will decrease again \cite{Batrouni}.

\begin{figure}[h!]
 \includegraphics[width=0.45\textwidth]{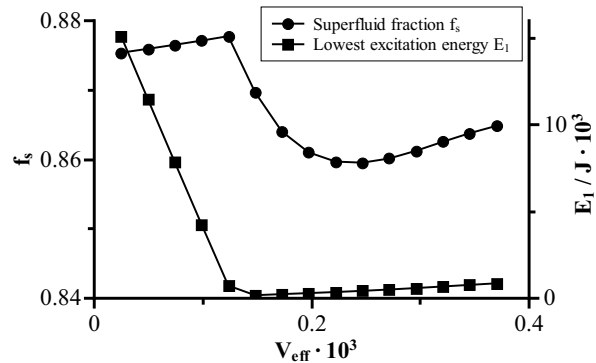}
 \caption{\label{fig:04}Superfluid fraction $f_{s}$ and  lowest excitation energy $E_{1}$ vs. $U$ at $\Delta = 1.13$.}
\end{figure}

Both the superfluid fraction and the lowest excitation energy show a sharp drop in the region where the interactions begin to fragment the condensate (Fig. \ref{fig:04}). The lowest excitation energy approaches zero where the energies for the single-peaked state and the fragmented state are equal. This effect results from the competition between the on-site energy and the kinetic and interaction energies. The single large peak minimizes the external potential energy while a fragmented condensate minimizes the kinetic and interaction energy. Above a critical interaction strength, the increased potential energy is compensated by the reduced kinetic and interaction energies. As each peak is exponentially localized, the superfluid fraction decreases at the critical interaction strength, but afterwards continues to increase as the interactions keep delocalizing the condensate.

\begin{figure}[h!]
 \includegraphics[width=0.45\textwidth]{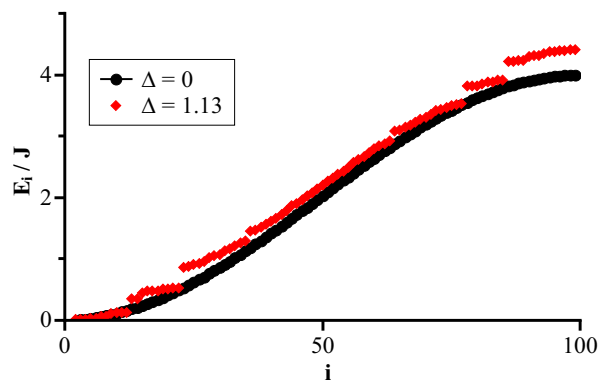}
 \caption{\label{fig:05}Band structure of the weakly interacting condensate
($V_{eff}= 10^{-8}$).}
\end{figure}

The excitation spectrum is shown in Fig. \ref{fig:05} for a weakly interacting condensate with and without disorder. In the absence of disorder, we obtain the band structure of a free particle in a sinusoidal potential. In the presence of disorder, the band shows gaps at fractions  corresponding to the lattice ratios $\alpha$ and $\beta$, demonstrating the quasi-periodicity of disorder. In the setup of \cite{Schulte}, for example, $\alpha$ is approximately $0.86$ and $\beta$ is approximately $0.78$. We find gaps, for example, at indices $86$ and $100 - 86 = 14$ due to $\alpha$, at $78$ and $100 - 78 = 22$ due to $\beta$ and at $35$ due to a beating, where $0.35 = 2 (\alpha^{-1}+(1-\beta)^{-1})^{-1}$. Our simulations show that $V_{eff}$ has no qualitative effect on the band structure. We conclude that the excitation spectrum is primarily governed by single-particle effects.

\subsection{Harmonic Trap}

\begin{figure}[h!]
 \centering{
 \subfloat[$\Delta = 0$]{
  \includegraphics[width=0.2\textwidth]{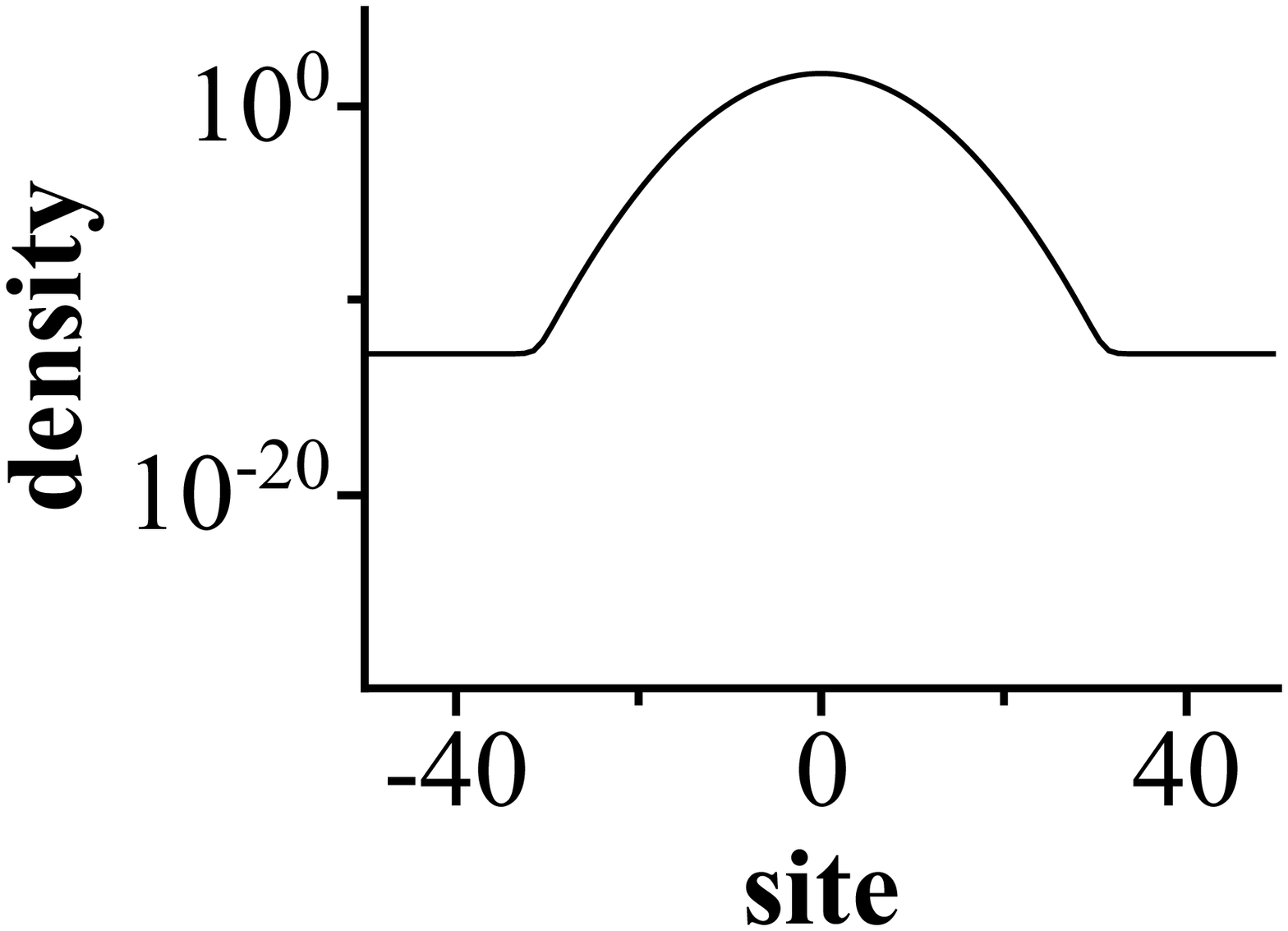}}
 \subfloat[$\Delta = 0.68$]{
  \includegraphics[width=0.2\textwidth]{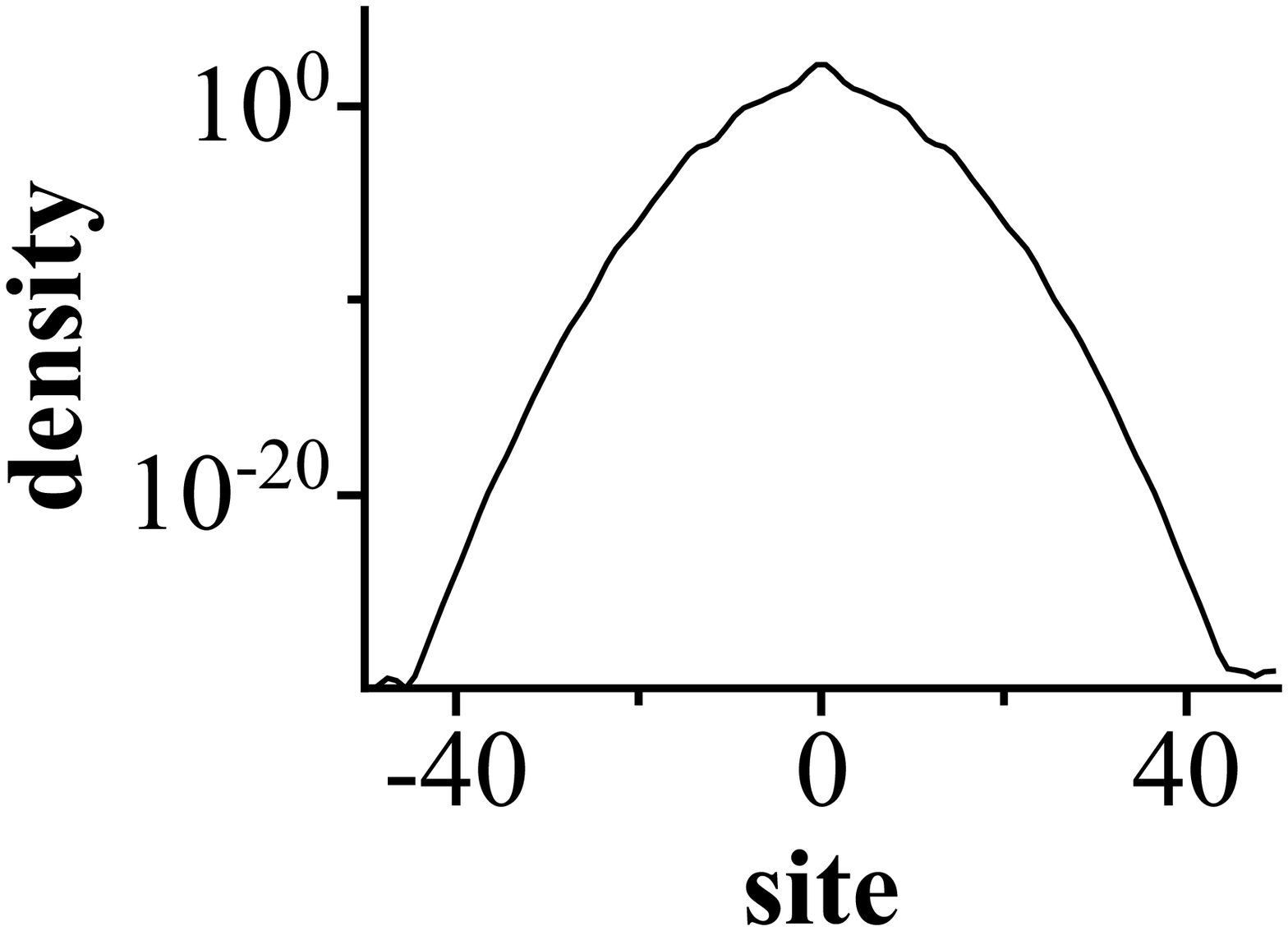}}

 \subfloat[$\Delta = 1.13$]{
  \includegraphics[width=0.2\textwidth]{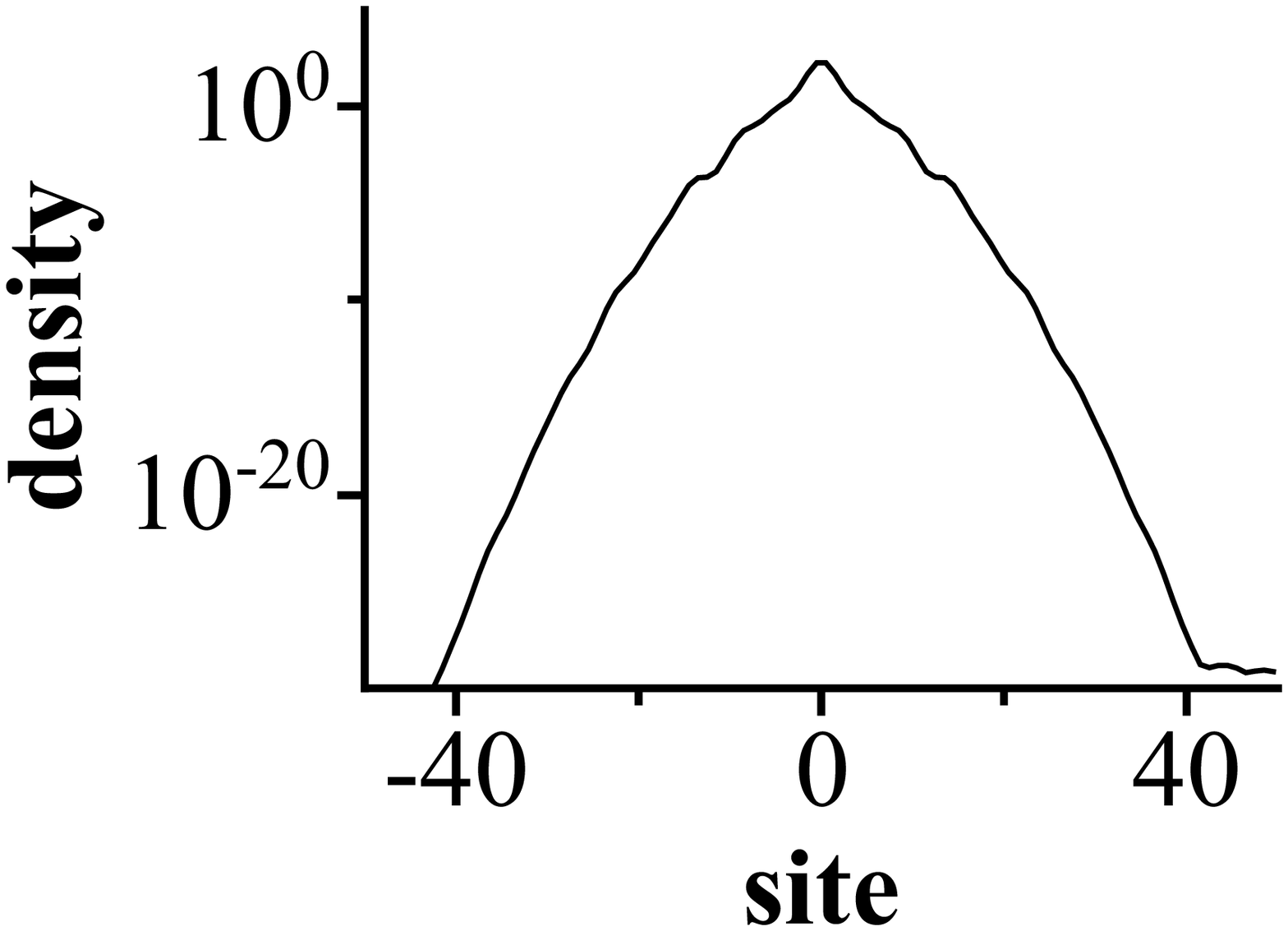}}
 \subfloat[$\Delta = 1.13$]{
  \includegraphics[width=0.2\textwidth]{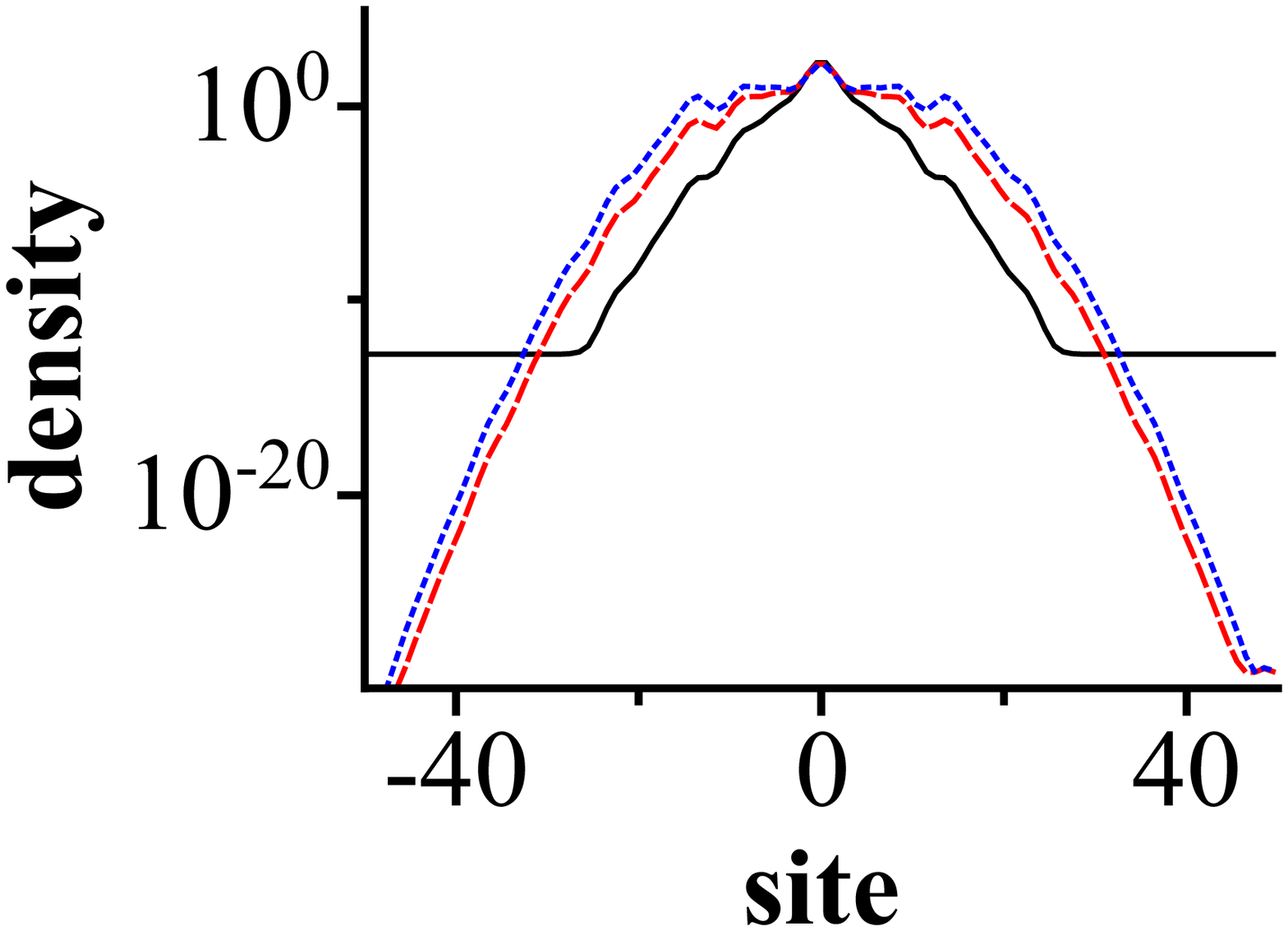}}
  }
 \caption{\label{fig:06}Condensate density (logarithmic) at $f_{t} = 10$Hz for various parameters. \mbox{(solid black) $V_{eff} = 10^{-8}$.} \mbox{(dashed red) $V_{eff} = 3.7\e{-4}$.} \mbox{(dotted blue) $V_{eff} = 6.2\e{-4}$.}}
\end{figure}

\begin{figure}[h!]
 \includegraphics[width=0.45\textwidth]{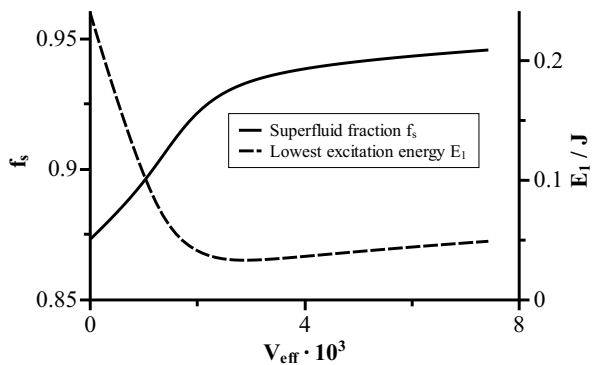}
 \caption{\label{fig:07}Superfluid fraction and lowest excitation energy vs. interaction strength.}
\end{figure}

The general observations of the previous section remain valid when a harmonic trapping potential is present, with the main difference being the additional confinement due to the trap. In this case, only in the logarithmic plots is it possible to distinguish between localization due to the trap, which is Gaussian, and localization due to Anderson localization, which is exponential (Fig. \ref{fig:06}). We also note that the trap inhibits the fragmentation of the condensate. This can be seen most clearly in the graphs for the superfluid fraction and the lowest excitation energy (Fig. \ref{fig:07}). In contrast to the homogeneous case, no sharp drops occur and the lowest excitation energy does not approach zero. We conclude that no sudden crossing from a single-peak state to a multiple-peak state occurs. Instead, side peaks arise continuously over a wide range of interaction strengths.


\subsection{Finite Temperature}

We now discuss the situation for finite temperature. For the mean-field approximation to remain valid, we must stay well below the critical temperature where the non-condensate density is low. For our simulations, we take the setup from the previous section and vary the temperature to $10$nK.

The effect of the finite temperature is to drive atoms out of the condensate into the thermal cloud. Thus the condensate fraction $f_{c}$ is lowered and the non-condensate density $\tilde{n}$ is increased. Because Anderson localization is a single-particle effect that does not depend on the total number of atoms in the condensate, we still see a Gaussian peak for weak disorder and exponential localization for strong disorder. The thermal cloud, while being slightly affected by the disorder potential, does not become localized. This is because the thermal cloud is incoherent and therefore behaves more like a classical gas.

\begin{figure}[h!]
 \includegraphics[width=0.45\textwidth]{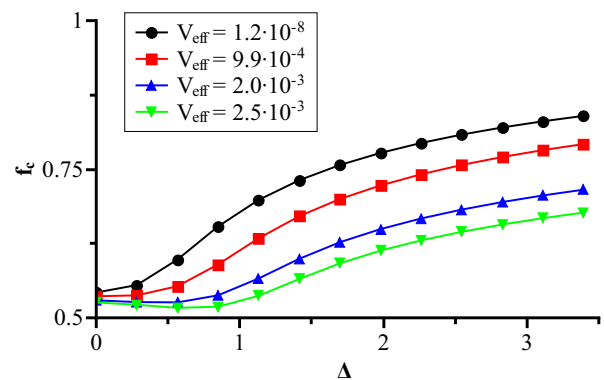}
 \caption{\label{fig:08}Condensate fraction vs. interaction strength at $T = 10$nK.}
\end{figure}

\begin{figure}[h!]
 \includegraphics[width=0.45\textwidth]{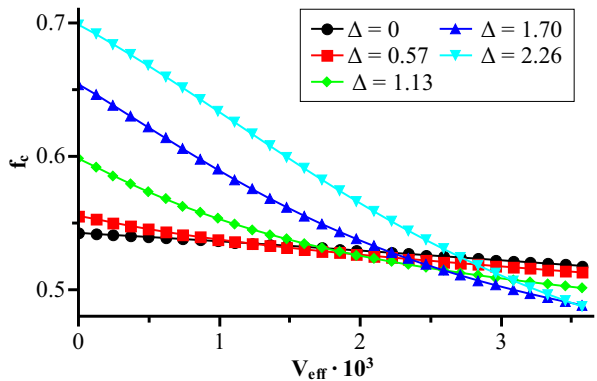}
 \caption{\label{fig:09}Condensate fraction vs. disorder strength at $T = 10$nK and $V_{eff}=10^{-8}$.}
\end{figure}

The interplay between disorder and interactions is more subtle at finite temperature. We first observe that increasing the interaction strength does not fragment the condensate. Instead, atoms are driven into the thermal cloud (Fig. \ref{fig:08}). This is energetically advantageous as the thermal cloud  has lower interaction energy due to its reduced density. We further see that disorder increases the condensate fraction for weak interactions (Fig. \ref{fig:09}) and decreases it for strong interactions. The former is a consequence of micro-trapping. The disordered potential contains microscopic wells whose harmonic trapping frequencies are higher than the $10$Hz of the external trap. This increases the critical temperature $T_{C}$, because $T_{C} \sim \omega_{t}$ for a BEC in a harmonic trap of trapping frequency $\omega_t$. Consequently, the ratio $T/T_{C}$ is decreased and the condensate fraction rises. For strong interactions, in contrast, the localizing effect of disorder
increases the interaction energy, thereby driving more atoms into the thermal cloud.

The superfluid fraction follows the condensate fraction because only atoms in the condensate contribute to the superfluidity. Apart from this, it does not show any new behavior differing from that of the zero-temperature case.\\

\subsection{Excitation Energies}

We conclude the treatment of the harmonically confined Bose gas at finite temperature with a discussion of the excitation energies. Again, we look at the interplay of temperature, disorder and interactions. We expect the behavior of $E_{1}$ to correspond to the behavior of $f_{c}$. Lower excitation energies should result in a lower condensate fraction because less thermal energy is needed to populate low lying levels.

\begin{figure}[!]
 \includegraphics[width=0.45\textwidth]{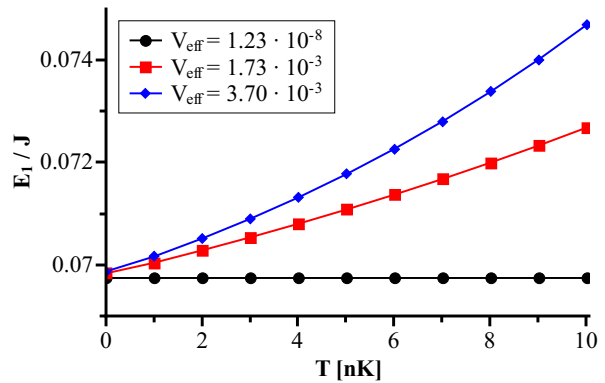}
 \caption{\label{fig:10}$E_{1}$ vs. $T$ for various interaction strengths at $\Delta = 0$.}
\end{figure}

First, we investigate the temperature dependence. Our simulations show that for weak interactions ($V_{eff} \sim 10^{-8}$), the temperature has essentially no effect on the band structure, regardless of the value of $\Delta$. This is not surprising as temperature does not alter the energy levels or states of a single particle system. In the presence of interactions, however, we see that increasing the temperature also increases the lowest excitation energy (Fig. \ref{fig:10}), and this effect becomes more pronounced as $V_{eff}$ increases. This can be  explained with the simultaneous decrease in the condensate fraction as temperature and interaction strength increase. Removing atoms from the condensate lowers the energy of the ground state via the interaction contribution $n_{c} V_{eff}$ and raises the energy of the excited states, thereby increasing the energy difference $E_{1}$ between the ground state  and the first excited state. It is worth noting that, contrary to our expectation, the condensate fraction does not increase with increasing excitation energy. This is attributed to the dependence of $E_{1}$ on $f_{c}$ in the presence of interactions, as discussed above.

\begin{figure}[!]
 \includegraphics[width=0.45\textwidth]{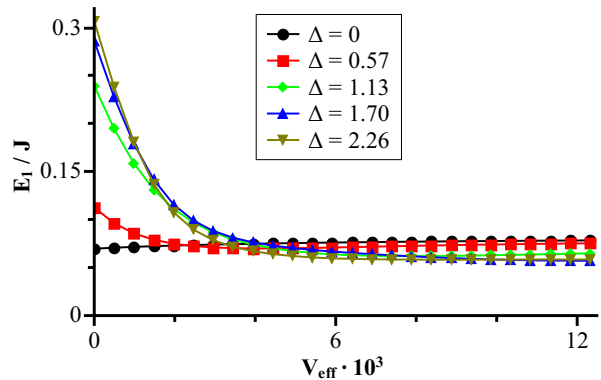}
 \caption{\label{fig:11}$E_{1}$ vs. $U$ for various disorder strengths at $T = 10$ nK.}
\end{figure}

Next, we turn to the interplay between disorder and interactions. Fig. \ref{fig:11} depicts the lowest excitation energy vs. $V_{eff}$. The graph shows that in the presence of disorder, increased interactions lower the excitation energy. At constant temperature, this allows more atoms to leave the condensate, in agreement with what we observe for $f_{c}$. The effect of disorder in the weakly interacting regime is to increase the excitation energy. For higher interaction strengths, this effect is reversed. This reflects the fact that more atoms leave the condensate when it becomes localized in the presence of strong interactions.\\

\section{Conclusion}

In this paper, we have discussed the interacting ultra-cold Bose gas in incommensurate optical lattices and highlighted the important role of 
disorder, interactions and temperature in the quest for Anderson localization. Our simulations verify that Anderson localization occurs 
and elucidate how finite temperature, a harmonic trapping potential and repulsive interactions affect the condensate.

\section{Acknowledgments}

This work was supported by contract SFB/TR 12 of the German Research Foundation and through the IB BMBF (Project NZL 07/006) 
and by the New Zealand Foundation for Research, Science and Technology through contract NERF-UOOX0703:Quantum Technologies 
and the New Zealand International Science and Technology Linkages Fund.


\end{document}